\documentclass[twocolumn,preprintnumbers,floatfix,prb,superscriptaddress]{revtex4}
\usepackage{graphicx}
\usepackage{dcolumn} 
\usepackage{bm}
\usepackage{epsfig}
\usepackage{longtable}
\begin{document} 

\title{Electron doped layered nickelates: spanning the phase diagram of the cuprates}

\author{A. S. Botana}
\affiliation{Materials Science Division, Argonne National Laboratory, Argonne, Illinois 60439, USA}
\author{V. Pardo}
\affiliation{Departamento de Fisica Aplicada, Universidade de Santiago de Compostela, E-15782 Santiago de Compostela, Spain}
\affiliation{Instituto de Investigacions Tecnoloxicas, Universidade de Santiago de Compostela, E-15782 Santiago de Compostela, Spain}
\author{M. R. Norman}
\email{norman@anl.gov}
\affiliation{Materials Science Division, Argonne National Laboratory, Argonne, Illinois 60439, USA}
\date{\today}

\begin{abstract}
Pr$_4$Ni$_3$O$_8$ is an overdoped analog of hole-doped layered cuprates. Here we show via \textit{ab initio} calculations that Ce-doped Pr$_4$Ni$_3$O$_8$ (Pr$_3$CeNi$_3$O$_8$) has the same electronic structure as the antiferromagnetic insulating phase of parent cuprates. We find that substantial Ce-doping should be thermodynamically stable and that other 4+ cations
 would yield a similar antiferromagnetic insulating state, arguing this configuration is robust for layered nickelates of low enough valence. The analogies with cuprates at different $d$ fillings suggest that intermediate Ce-doping concentrations near
 1/8 should be an appropriate place to search for superconductivity in these low-valence Ni oxides. 

\end{abstract}

\maketitle

\begin{figure}
\includegraphics[width=\columnwidth,draft=false]{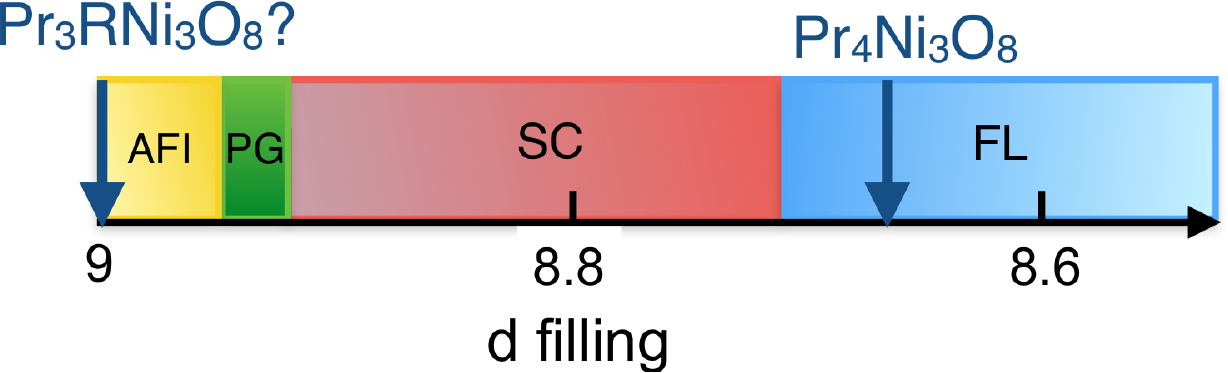}
\caption{ A schematic diagram representing some of the phases of the cuprates as a function of nominal filling of the $d$ levels in the hole-doped region: antiferromagnetic insulating (AFI), pseudogap (PG), superconducting (SC), and Fermi liquid (FL). Pr$_4$Ni$_3$O$_8$  and Pr$_3$RNi$_3$O$_8$ (R representing a 4+ cation, namely Ce) are represented at the appropriate $d$ fillings.}\label{fig1}
\end{figure}

Cuprate-like electronic structures have been pursued in other oxides since soon after high temperature superconductors were discovered.\cite{norman, pickett_cuprates, cuprates_r, cuprates_dagotto} Cuprates are layered materials with an underlying CuO$_2$ square lattice. The parent compounds (Cu$^{2+}$: d$^9$, S= 1/2) have 
d$_{x^2-y^2}$ as the active orbitals, with strong antiferromagnetic correlations, as well as a substantial hybridization of Cu-$d$ and O-$p$ states. Upon doping the CuO$_2$ planes away from a stoichiometric Cu$^{2+}$ oxidation state, the antiferromagnetic state is suppressed, then superconductivity appears, and at higher dopings a Fermi liquid phase arises.\cite{pickett_cuprates} A schematic phase diagram with respect to filling of the $d$ levels for hole doping is shown in Fig.~\ref{fig1}. 

Given the proximity of Ni and Cu in the periodic table, an interesting conjecture is whether these phenomena could also occur in nickelates,
since if so, this could tell us a lot about the nature of high temperature superconductivity.
LaNiO$_3$-based heterostructures \cite{FirstHeterostructure,Heterostructure_review} and doped single layer nickelates (La$_{2-x}$Sr$_x$NiO$_4$) \cite{single_layer,single_layer_2} have been intensively studied, but the challenge is finding a nickelate that has a similar electron count (Ni$^{1+}$ and Cu$^{2+}$ being isoelectronic) with Ni in a square planar environment.

In this context, the recently synthesized low valence layered nickelates (R$_4$Ni$_3$O$_8$, R438, with R= La or Pr) with an average Ni valence of +1.33, are promising candidates to investigate.\cite{la438_1, la438_2, Pardo_Quantum, Poltavets_2010PRL, LACORRE_1992, AprobertsWarren_NMRLa4Ni3O8, Sarkar_La3Ni2O6andLa4Ni3O8, scalettar}
Particularly, Pr438, a 1/3 self hole doped (relative to d$^9$) line compound, stands among the closest bulk analogs to high-T$_c$ cuprates. It shows metallic behavior and a large orbital polarization with holes in the planar d$_{x^2-y^2}$ states, as well as a high degree of O-$p$--Ni-$d$ hybridization.\cite{nat_phys} Fig.~\ref{fig1} illustrates where metallic Pr438 lies within the cuprate phase diagram, sitting in the overdoped Fermi liquid region.

Electron doping this new phase is an obvious strategy to explore the analogy with cuprates. If we consider as a starting point that the physics of high-T$_c$ superconductivity in cuprates is brought about by doping a 2D Mott insulator with strong  antiferromagnetic correlations, a natural question is whether this type of ``parent phase'' arises at $d^9$ filling for electron-doped Pr438.\cite{review} As pointed out almost 20 years ago by Anisimov \textit{et al.},\cite{anisimov} ``only if the Ni ions are
forced into a planar coordination with the O ions can a S=1/2 magnetic insulator be realized with the difficult
Ni$^{1+}$ oxidation state".
Here, we show using DFT-based calculations that doping Pr$_4$Ni$_3$O$_8$ with a 4+ ion allows going from a Fermi liquid phase at a nominal d$^{8.67}$ filling to a parent antiferromagnetic insulating phase at d$^9$ filling, effectively spanning the hole-doped cuprate phase diagram (see Fig.\ref{fig1}).

Our electronic structure calculations were performed within density functional
theory using the all-electron, full potential code {\sc wien2k} based on the augmented plane wave plus local orbitals (APW+lo) basis set.\cite{wien2k, sjo} For the structural relaxations we have used the Perdew-Burke-Ernzerhof version of the generalized gradient approximation (GGA).\cite{pbe}  Volume and $c/a$ optimizations, as well as a full relaxation of all the internal atomic coordinates, were carried out. To deal with  strong correlation effects,
we apply the LDA+$U$ scheme that incorporates an on-site Coulomb repulsion $U$ and Hund's rule coupling strength $J_H$.\cite{sic}
We use typical values: $U$= 5 eV for Ni-$3d$, $U$= 8 eV for Pr-$4f$, and $U$= 5 eV for Ce-$4f$. $J_H$ has been set to 0.7 eV for Ni-$d$ and 1 eV for Pr and Ce-$f$. 
Experience with the cuprates and other transition metal oxides
shows LDA+$U$ to be a reliable method to predict the
electronic structure. For example, the antiferromagnetic insulating
state of the stoichiometric cuprates is
correctly reproduced.\cite{u_cuprates, u_cuprates_2, cuprates_gap}

\begin{figure}
\includegraphics[width=\columnwidth,draft=false]{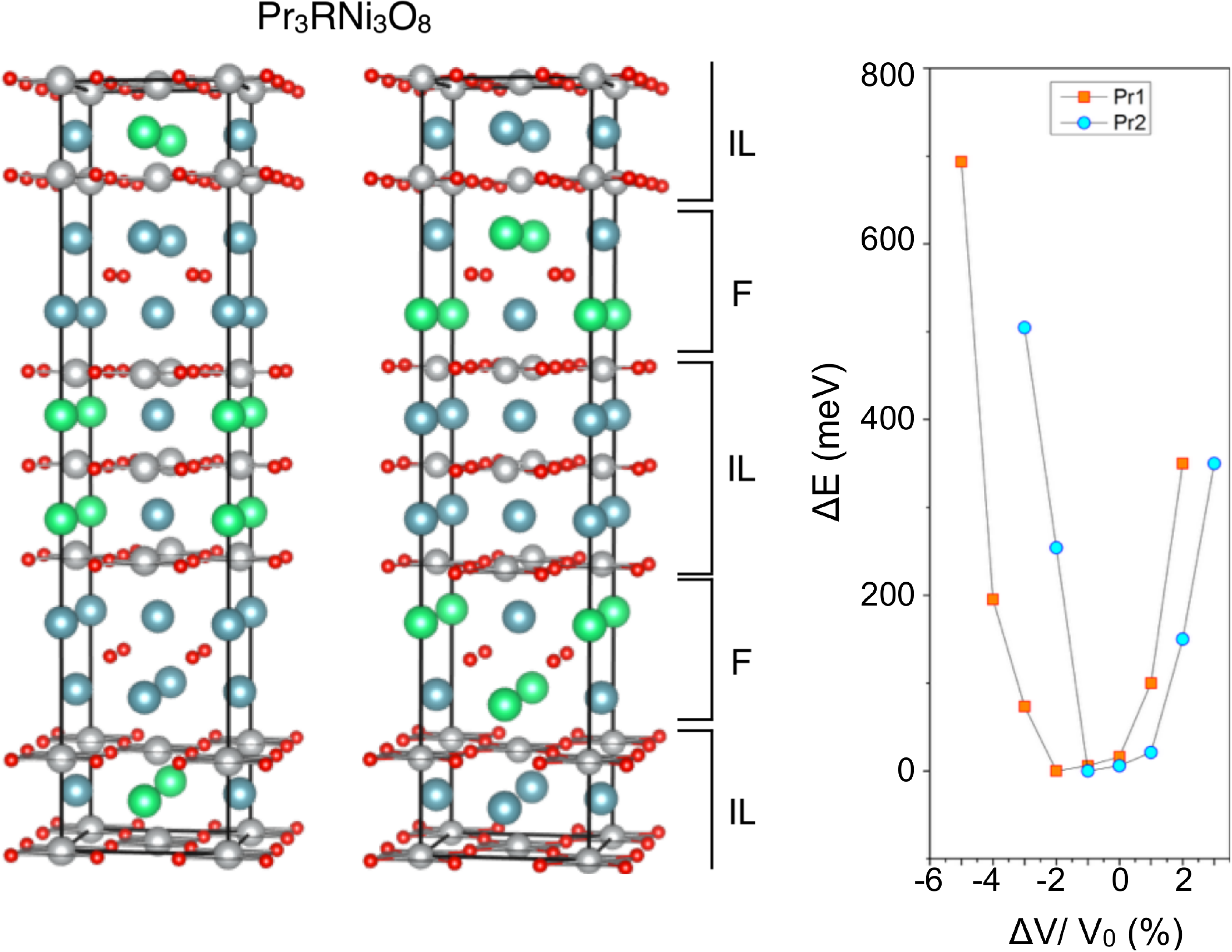}
\caption{Left panels. Structure of the $\sqrt2$$a$ $\times$ $\sqrt2$$a$ $\times$ $c$ cells for Pr$_{3}$CeNi$_3$O$_8$ with Ce substitution on both Pr1 and Pr2 sites (infinite layer, IL, and fluorite, F, blocks, respectively).  Ce atoms are in green, Pr atoms in blue, Ni atoms in gray and O atoms in red. Right panel: Energy versus volume curves for both structures obtained within GGA-PBE. V$_0$ represents the experimental volume of Pr438.}\label{struct}
\end{figure}

Pr438 crystallizes in a tetragonal unit cell,\cite{nat_phys} with space
group I4/mmm (no.~139) and lattice parameters a=
3.9347 \AA , c= 25.4850 \AA . The structure, pictured in
Fig.~\ref{struct}, consists of three NiO$_2$ infinite-layer (IL) planes separated
by layers of Pr ions. On either
side of this trilayer lies a fluorite (F) Pr-O$_2$-Pr block. With an average Ni valence of ${1.33+}$, Pr438 is isoelectronic with (hole) overdoped cuprates involving a mixture of Ni$^{1+}$ and Ni$^{2+}$ ions. The electronic structure of Pr438 is found to be metallic, consistent with a 2/3-filled Ni-d$_{x^2-y^2}$ band crossing the Fermi level hybridized with O-$p$ states as described in Ref.~\onlinecite{nat_phys}.

By substituting trivalent Pr with a tetravalent ion in Pr438, the electron count on the Ni-O layer can be changed, moving the material towards the left of the phase diagram shown in Fig.~\ref{fig1}.  In principle, the ionic count required for reaching the parent insulating state (with Ni$^{1+}$: $d^9$) would need 25\% substitution on the Pr site with a 4+ cation (Pr$_3$RNi$_3$O$_8$, R= Ce, Th). 
However, the low oxidation state Ni$^{1+}$ is
hard to stabilize. The bulk synthesis of the Ni$^{1+}$ infinite-layer
compound LaNiO$_2$ is difficult and because of these materials issues, it is unclear whether an insulating state is achieved.\cite{lanio2_1, lanio2_2, lanio2_pickett}
LDA+$U$ calculations give a stable solution with antiferromagnetic ordering of S=1/2 ions,\cite{anisimov} but experimentally there is no evidence for long range order. In principle, the on-site energy difference between the $p$ and $d$ levels is smaller for Cu$^{2+}$ than for Ni$^{1+}$, favoring larger hybridizations in the former. However, we will see that substantial oxygen hybridization can occur for the layered nickelates studied here.

\begin{table}
\caption{Nearest neighbor Pr-O, Ce-O and Ni-O distances for Pr$_3$CeNi$_3$O$_8$ in \AA~ for Ce substitution on the Pr1 and Pr2 sites. The Pr/Ce environment is square antiprismatic with eight oxygen ions as nearest neighbors. The R ion is positioned asymmetrically inside the polyhedra. For this reason, two different R-O distances are listed in the table. In undoped Pr438 the Pr-O distances are (in \AA) 2.48, 2.59 for Pr1 (IL), and 2.32, 2.74 for Pr2 (F). The Ni is in a square planar environment of oxygens. For undoped Pr438 the Ni-O distance is 1.97 \AA. }\label{table1}
\begin{center}
\begin{tabular}{ c c c  c c   c}
\hline
     \hline
    & Pr$_3$CeNi$_3$O$_8$(IL) &&Pr$_3$CeNi$_3$O$_8$(F) & \\    
\hline
$Ce-IL$&    2.39, 2.54 &  $Ce-F$&    2.30, 2.63 & \\
$Pr-IL$&    2.46, 2.59&  $Pr-F$&    2.40, 2.63 & \\
$Pr-F$&    2.33, 2.74 & $Pr-IL$&      2.53 2.59 &\\ 
$Pr-F$&    2.32, 2.79&  $Pr-IL$&     2.55, 2.58& \\
$Ni$&   1.96  &   $Ni$&   1.96  &\\
\hline
\hline
\end{tabular}
\end{center} 
\end{table}

In Pr438 there are two different Pr sites: Pr1 within the NiO$_2$ trilayer and Pr2 within the fluorite blocks that separate NiO$_2$ trilayers (see Fig.~\ref{struct}). Both types of substitutions were tried, giving rise to a very similar outcome for the electronic structure (which will be described in detail below). Concerning the most appropriate 4+ dopant atom, Ce is the most obvious choice given its proximity to Pr in the periodic table and because Ce-doping has been realized in T$^{\prime}$-cuprates giving rise to electron-doped phases. For instance, T$^{\prime}$-Pr$_{2-x}$Ce$_x$CuO$_4$ shows superconductivity with an optimum doping level x= 0.135 and T$_c$= 27 K.\cite{ce_doping_nat, electron_doped_cuprates, el_doped_cuprates_2} In T$^{\prime}$-Nd$_2$CuO$_4$ superconductivity develops by doping with both Ce and Th.\cite{ndcu_ce_doped_1,el_doped_cuprates}

To test the possibility of in-plane checkerboard magnetic (AFM) ordering in Pr$_3$CeNi$_3$O$_8$, a $\sqrt2$$a$ $\times$ $\sqrt2$$a$ $\times$ $c$ cell was used. After the substitution of Pr by Ce in a 3:1 ratio, the optimized volume results in a reduction with respect to that of Pr438 of 1 to 2 \% (see Fig.~\ref{struct}) -- the ionic radius of Ce$^{4+}$ in an 8-fold coordination is 0.97 \AA ~vs the ionic radius of Pr$^{3+}$ of 1.13 \AA, so a decrease in lattice constants can be anticipated.\cite{shannon} Small distortions of the Ni-O and Pr/Ce-O distances take place after structural relaxation as shown in Table \ref{table1}.  Ce-substitution within the F block is energetically favorable by 96 meV/unit cell with respect to substitution in the IL block. The standard enthalpy
change $\Delta$H can be obtained at T= 0 from the calculated total energies: E(Pr$_3$CeNi$_3$O$_8$)+ $\frac{3}{4}
$E(O$_2$)- $\frac{3}{2}$E(Pr$_2$O$_3$)- E(CeO$_2$)- 3E(NiO) giving a value of -814 $kJ/mol$.

\begin{figure}
\includegraphics[width=\columnwidth,draft=false]{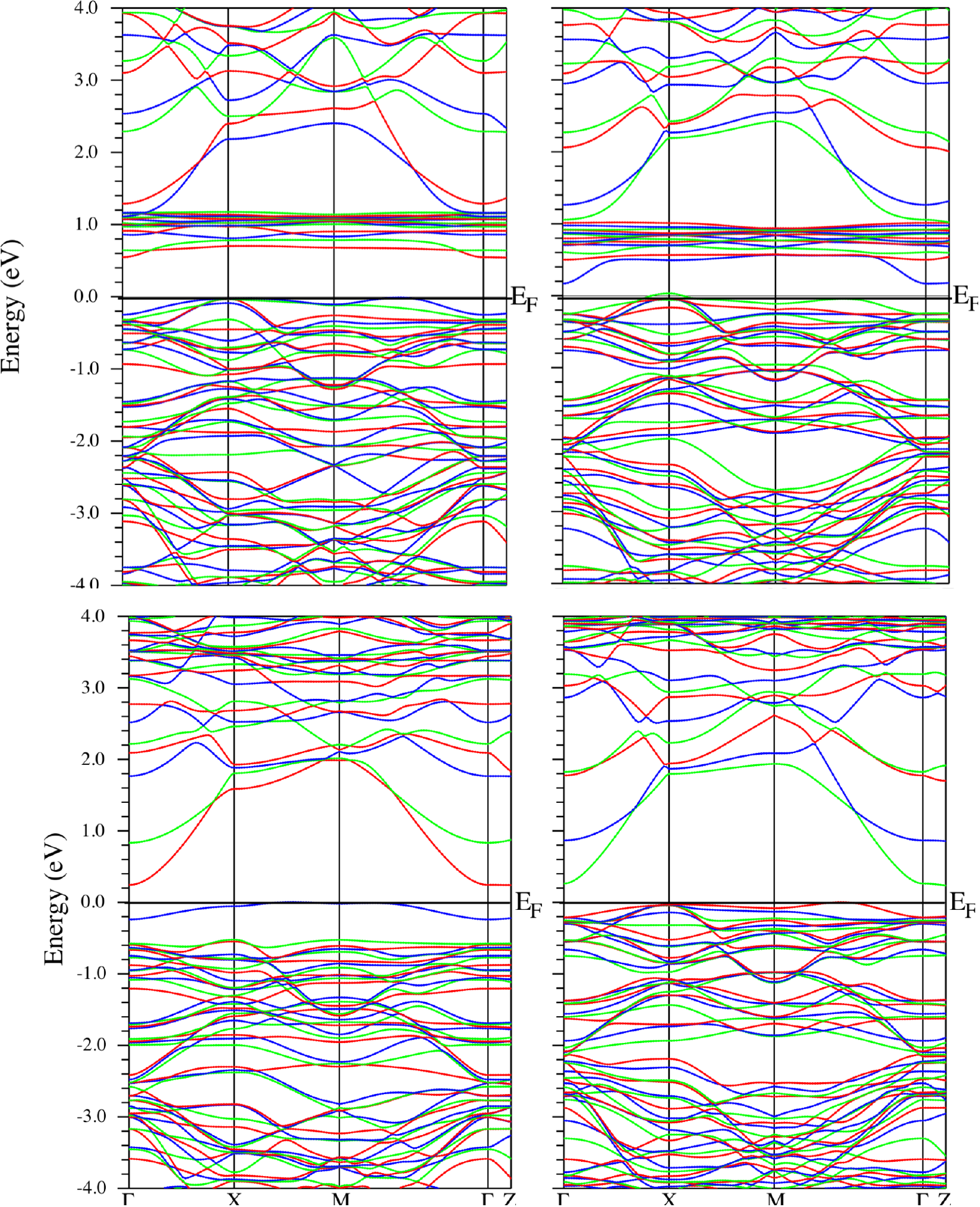}
\caption{Band structures for  Pr$_{3}$RNi$_3$O$_8$ within LDA+$U$ (R= Ce, top panels; R= Th, bottom panels). R$^{4+}$ doping within the IL (left panels) and F blocks (right panels) with an in-plane AFM ordering of Ni ions. }\label{bs}
\end{figure}

\begin{figure}
\includegraphics[width=\columnwidth,draft=false]{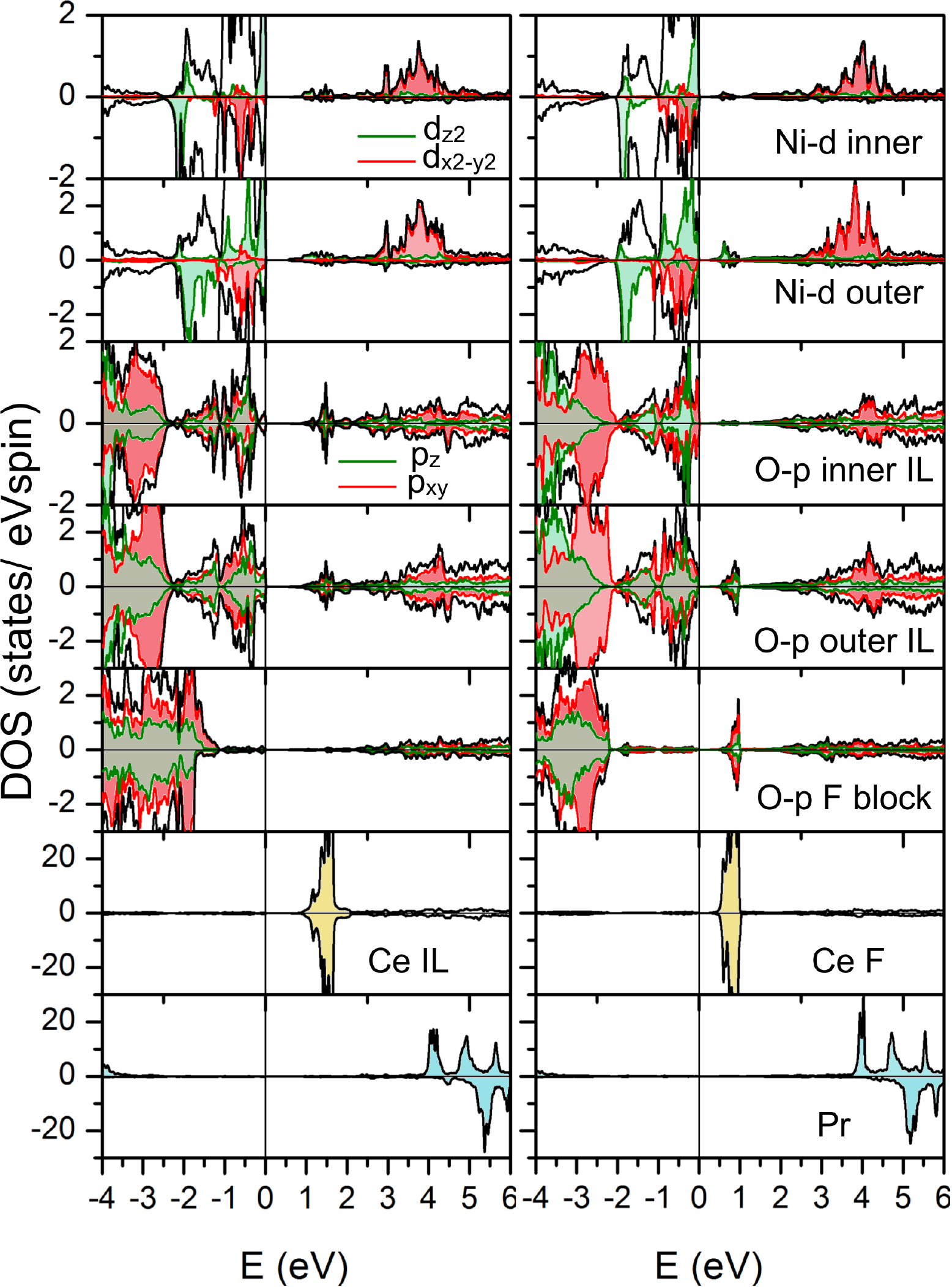}
\caption{Partial density of states for Pr$_{3}$CeNi$_3$O$_8$ within LDA+$U$ (top parts of panels are for spin up, bottom for spin down). Ce$^{4+}$ doping within the IL block (left) and within the fluorite block (right). We find that the Ce cation slightly bonds with neighboring O ions. The unoccupied Ni-$d$ bands lying at about 3 eV above the Fermi level have substantial O character, as a result of the strong $\sigma$-bond with neighboring O anions.}\label{dos}
\end{figure}

Within GGA, the band structure is metallic with an in-plane AFM coupling within the NiO$_2$ planes. Application of a Coulomb $U$ to the Ni-$d$, Pr-$f$ and Ce-$f$ states is enough to open a gap of $\sim$0.6 eV for doping in the IL block and $\sim$0.3 eV in the F block (see Fig.~\ref{bs}). The gap is smaller for substitution in the F block
due to the unoccupied Ni-$d$ states from the outer planes being closer in energy to the Ce-$f$ bands,
leading to a level repulsion effect (see Fig.~\ref{dos}).
The magnetic moments inside the Ni muffin-tin sphere are approximately $\pm$0.9 $\mu_B$  with checkerboard (AFM) ordering within the NiO$_2$ planes as well as between planes. All in all, our LDA+$U$ band structure, shown in Fig.~\ref{bs}, exhibits a correlated S=1/2 AFM insulator analog of the parent cuprate materials. Holes would hence go into the Ni-$d$ band of $x^2-y^2$ character that is highly hybridized with O-$p$ states (see below). Calculations confirm that this state comprised of AFM-ordered Ni layers is more stable than a ferromagnetic one by  a sizable 0.2 eV/Ni,
implying a superexchange interaction comparable to that of cuprates.

In general, Ce can be trivalent as well as tetravalent. Using the same structural parameters and $U$ values described above, as well as a checkerboard AFM configuration within the NiO$_2$ planes, a solution where Ce is 3+  was obtained. However, this state turns out be energetically unfavorable by 0.34 eV/unit cell with respect to that where Ce is tetravalent. The derived electronic structure is metallic with a magnetic moment of 1 $\mu_B$ developing on Ce. An in-trimer charge ordered solution results with AFM ordering in the NiO$_2$ planes: Ni$^{1+}$ in the outer layers, and Ni$^{2+}$ (high spin) in the inner layers. \cite{wu_co}

\begin{figure}
\includegraphics[width=\columnwidth,draft=false]{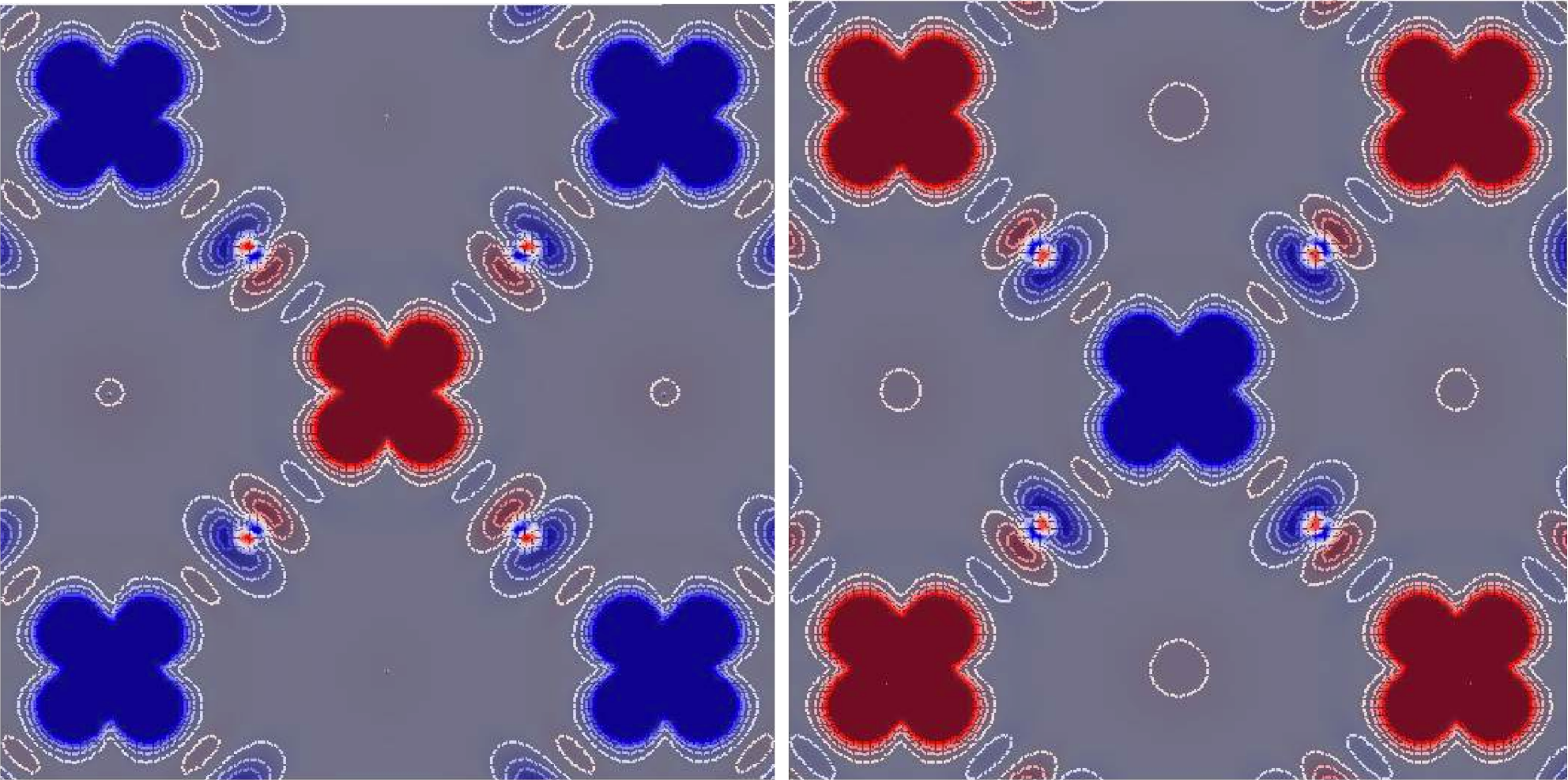}
\caption{Contour plot of the spin density within the NiO$_2$ layers (left outer, right inner layer) showing the checkerboard AFM ordering. The Ni-$d_{x^2-y^2}$ orbitals point directly to the $p$ orbitals of the neighboring oxygens. Different colors represent spin up and spin down.}\label{density}
\end{figure}

As can be observed in Fig.~\ref{dos}, for Ce$^{4+}$ substitution, the top of the valence band consists of a mixture of $d_{x^2-y^2}$ and $d_{z^2}$ states, whereas the Ni-$d$ unoccupied bands are $x^2-y^2$-only in character. At around 1 eV above the Fermi level lie the Ce-$f$ states whereas Pr-$f$ states appear mostly above $\sim$ 4 eV. It is clear from the density of states (DOS) that both inner and outer Ni$^{1+}$: $d^9$ (S=1/2) ions have a single hole in the minority-spin $d_{x^2-y^2}$-like band. Hybridization between the Ni-$d$ and O-$p$ states is clear in the DOS, being most noticeable for the planar O ions as a result of the strong $pd$$\sigma$ bond. The lobes of the $x^2-y^2$ orbital point directly to the $p$ orbital of the neighboring oxygen, forming a strong covalent bond with a large hopping integral $t_{pd}$= 0.3 eV. The strong directionality of the bond can be observed for Pr$_{3}$CeNi$_3$O$_8$ in Fig.~\ref{density}. Let us recall that  bond covalency sets the unusually high energy scale for the exchange interaction in cuprates. Small moments are induced on the oxygen ions as a consequence of hybridization along with low-site symmetry: $\mu_{O1}$= -0.05 $\mu_B$ (inner NiO$_2$ plane), $\mu_{O2}$=-0.07 $\mu_B$ (outer NiO$_2$ plane), $\mu_{O3}$= -0.02 $\mu_B$ (F block).
Importantly, the Ce cation bonds with neighboring O ions (as implied in the DOS, clearly reflecting the hybridization between O-$p$ and Ce-$f$ states). Previous results show that Ce substitution in R$_2$CuO$_4$ (R= Pr, Nd) enhances the phase stability given that tetravalent Ce has a firmer grip on surrounding O$^{2-}$ ions by stronger electrostatic force than trivalent rare earth ions.\cite{el_doped_cuprates, thermodyn_214_1, thermodyn_214_2}

To check the robustness of the obtained electronic structure, Th substitution on the Pr site was also tried as well as smaller 4+ cations like Zr and Te. For Th substitution, the same type of calculations give rise to a smaller volume reduction of only 0.5-1\% (the ionic radius of tetravalent Th$^{4+}$ in an 8-fold environment is 1.06 \AA, larger than that for Ce$^{4+}$).\cite{shannon} The electronic structure is very similar except for the $f$-states missing at the tail of the unoccupied Ni-$d$ states (see Fig.~\ref{bs}) arguing that the derived insulating AFM state is robust in Pr$_{3}$RNi$_3$O$_8$. Zr and Te are smaller ions \cite{shannon} so substitution with these gives rise to bigger distortions in the structure. However, the main features of the electronic structure obtained for Th doping remain for both Zr and Te doping.

If the likelihood of promoting superconductivity in these layered nickelates is to be analyzed, the similarities
with cuprates are important to characterize. The state we have obtained in Pr$_{3}$CeNi$_3$O$_8$ (for a Ni$^{1+}$ oxidation state) has insulating,
AFM planar order in common with the cuprates, as
well as obvious similarities in structure with common MO$_2$ (M= Ni, Cu) square lattice planes.
Even though we are dealing with hypothetical structures, our calculations show that substantial Ce-doping should be thermodynamically stable and that several other 4+ cations would yield a very similar antiferromagnetic insulating solution, suggesting this configuration is robust in layered nickelates of low enough valence.
Intermediate Ce-doping concentrations near 1/8 should be the appropriate place to search for superconductivity in these low-valence Ni oxides. Our results show that planar nickelates directly analogous to the high-T$_c$ superconducting cuprates are a realistic possibility,
especially given their similar superexchange interaction.

The authors acknowledge fruitful discussions with Peter Blaha, Warren Pickett and John Mitchell. ASB and MRN were supported by the Center for Emergent Superconductivity, an Energy Frontier Research Center funded by the US DOE, Office of Science, under Award No.~DE-AC0298CH1088. VP thanks MINECO of Spain for financial support through project No.~MAT2016-80762-R. We acknowledge the computing resources provided on Blues, the high-performance computing clusters operated by the Laboratory Computing Resource Center at Argonne National Laboratory.

\end{document}